\documentclass[10pt]{article} 
\usepackage[a4paper,total={6.5in, 9in}]{geometry}

\usepackage{glossaries}
\newacronym{ap}{AP}{Average Precision}
\newacronym{auc}{AUROC}{Area Under the Receiver Operating Curve}
\newacronym{ace}{ACE}{Automatic Concept Extraction}
\newacronym{ecg}{ECG}{electro cardiogram}
\newacronym{nmf}{NMF}{Non-negative Matrix Factorization}
\newacronym{pca}{PCA}{Principal Component Analysis}
\newacronym{ai}{AI}{Artificial Intelligence}
\newacronym{dl}{DL}{Deep Learning}
\newacronym{lof}{LOF}{Local Outlier Factor}
\newacronym{dnn}{DNN}{Deep Neural Network}
\newacronym{lrp}{LRP}{Layer-wise Relevance Propagation}
\newacronym{fda}{FDA}{Fisher Discriminant Analysis}
\newacronym{lsb}{LSB}{least significant bit}
\newacronym{xai}{XAI}{eXplainable Artificial Intelligence}
\newacronym{crp}{CRP}{Concept Relevance Propagation}
\newacronym{amax}{ActMax}{Activation Maximization}
\newacronym{rmax}{RelMax}{Relevance Maximization}
\newacronym{dora}{DORA}{Data-agnOstic Representation Analysis}
\newacronym{aoc}{AOC}{Area Over Curve}
\newacronym{conv}{Conv}{convolutional}
\newacronym{svm}{SVM}{Support Vector Machine}
\newacronym{gmm}{GMM}{Gaussian Mixture Model}
\newacronym{roi}{ROI}{Region of Interest}
\newacronym{lcrp}{L-CRP}{CRP for Localization Models}
\newacronym{leace}{LEACE}{LEAst-squares Concept Erasure}
\newacronym{rrr}{RRR}{Right for the Right Reason}
\newacronym{cdep}{CDEP}{Contextual Decomposition Explanation Penalization}
\newacronym{clarc}{ClArC}{Class Artifact Compensation}
\newacronym{aclarc}{\mbox{A-ClArC}}{Augmentive ClArC}
\newacronym{pclarc}{\mbox{P-ClArC}}{Projective ClArC}
\newacronym{rpclarc}{\mbox{rP-ClArC}}{reactive P-ClArC}
\newacronym{rrclarc}{RR-ClArC}{Right Reason ClArC}
\newacronym{ml}{ML}{Machine Learning}
\newacronym{iou}{IoU}{Intersection over Union}
\newacronym{cse}{CSE}{complete skin examination}
\newacronym{cav}{CAV}{Concept Activation Vector}
\newacronym{cam}{CAM}{Class Activation Maps}
\newacronym{car}{CAR}{Concept Activation Region}
\newacronym{pcx}{PCX}{Prototypical Concept-based eXplanations}
\newacronym{tcav}{TCAV}{Testing with CAV}
\newacronym{spray}{SpRAy}{Spectral Relevance Analysis}
\newacronym{svd}{SVD}{Singular Value Decomposition}
\newacronym{nams}{n-AMS}{natural Activation-Maximization signals}
\newacronym{sae}{SAE}{Sparse Autoencoder}
\newacronym{shap}{SHAP}{SHapley Additive exPlanations}
\newacronym{iterrev}{IterRev}{Iteratively Revealing and Revising Spurious Model Behavior}
\newacronym{r2r}{R2R}{Reveal to Revise}
\newacronym{xil}{XIL}{eXplanatory Interactive Learning}
\newacronym{sem}{SEM}{Standard Error of the Mean}
\newacronym{se}{SE}{Standard Error}
\newacronym{lvh}{LVH}{left ventricular hypertrophy}
\newacronym{tsne}{t-SNE}{t-Distributed Stochastic Neighbor Embedding}
\newacronym{umap}{UMAP}{Uniform Manifold Approximation and Projection}
\newacronym{vit}{ViT}{Vision Transformer}
\newacronym{slic}{SLIC}{simple linear iterative clustering}
\usepackage{amsmath,amsfonts,bm}

\def\eqref#1{equation~\ref{#1}}
\def\1{\bm{1}}

\DeclareMathAlphabet{\mathsfit}{\encodingdefault}{\sfdefault}{m}{sl}
\SetMathAlphabet{\mathsfit}{bold}{\encodingdefault}{\sfdefault}{bx}{n}

\makeatletter
\DeclareRobustCommand\onedot{\futurelet\@let@token\@onedot}
\def\@onedot{\ifx\@let@token.\else.\null\fi\xspace}

\usepackage[numbers]{natbib}
\usepackage{multibib}
\usepackage{hyperref}
\usepackage{url}
\usepackage{graphicx}
\usepackage{duckuments}
\usepackage{lmodern}
\usepackage{anyfontsize}
\usepackage{bookmark}
\usepackage{xspace}
\usepackage{multirow} 
\usepackage{booktabs} 
\usepackage{tikz}
\usepackage{multicol}

\definecolor{myblue}{HTML}{1F77B4}
\definecolor{mypurple}{HTML}{8172B3}
\definecolor{myorange}{HTML}{FF7F0E}
\definecolor{myred}{HTML}{C44E52}
\definecolor{mygrey}{HTML}{808080}
\definecolor{mygreen}{HTML}{2CA02C}

\definecolor{myredcircle}{HTML}{a12828} 
\definecolor{mybluecircle}{HTML}{5880A0}

\usepackage{hyperref}
\glsdisablehyper

\title{\textbf{An Analysis of the Interdependence Between Peanut and Other Agricultural Commodities in China's Futures Market}}
\author{
    Suke Li\textsuperscript{\rm 1},
}
\date{\small
    \textsuperscript{\rm 1}School of Software and Microelectronics, Peking University, Beijing, China\\
   \texttt{lisuke@ss.pku.edu.cn}\\
}

\begin{document}
\maketitle

% \icmltitle{Navigating Neural Space: Overcoming Directional Shifts via\\Signal-based Concept Activation Vectors}

% something with revisiting

% \icmltitle{Navigating Neural Space: Understanding the Significance of Directional Shifts in Concept Activation Vectors}




\begin{abstract}
This study analyzes historical data from five agricultural commodities in the Chinese futures market to explore the correlation, cointegration, and Granger causality between Peanut futures and related futures. Multivariate linear regression models are constructed for prices and logarithmic returns, while dynamic relationships are examined using VAR and DCC-EGARCH models. The results reveal a significant dynamic linkage between Peanut and Soybean Oil futures through DCC-EGARCH, whereas the VAR model suggests limited influence from other futures. Additionally, the application of MLP, CNN, and LSTM neural networks for price prediction highlights the critical role of time step configurations in forecasting accuracy. These findings provide valuable insights into the interconnectedness of agricultural futures markets and the efficacy of advanced modeling techniques in financial analysis.

\end{abstract}
% \keywords{Explainable Artificial Intelligence, Interpretability, Spurious Correlations, Bias Mitigation, Data Annotation}

\section{Introduction}

Peanut futures were officially launched on the Zhengzhou Commodity Exchange (ZCE) on February 1, 2021. Peanut futures are the first futures contract specifically designed for peanuts in China. They play a key role in facilitating price discovery and risk management within the peanut industry, and have been received widespread attention. Each Peanut futures contract represents a trading unit of 5 metric tons, quoted in Chinese Yuan (CNY) per ton, with a minimum price fluctuation of 2 CNY per ton. The daily price movement is restricted to ±4\% of the previous trading day's settlement price, in accordance with the ZCE's risk control regulations. A minimum trading margin of 5\% of the contract value is required. The contract months include January, March, April, May, October, November, and December, with trading hours from 9:00 AM to 11:30 AM and 1:30 PM to 3:00 PM, as well as other times specified by the exchange\citep{peanut2021future}.

In China, peanuts are primarily utilized as a raw material for the extraction of peanut oil, which holds a significant position in Chinese culinary practices as a key cooking oil. Additionally, peanuts are processed into a variety of food products. However, apart from peanut oil, other edible oils such as rapeseed oil, soybean oil, and palm oil are also widely used in cooking. In the absence of a distinct consumer preference, peanut oil can be substituted by these alternative oils. Consequently, this study hypothesizes that certain other futures commodities may exert a substantial influence on the price volatility of Peanut futures. Furthermore, the residual by-products from peanut oil extraction, often used as animal feed, establish a substitutable relationship with soybean meal, a by-product of soybean oil production. Therefore, this paper also examines the relationship between Soybean Meal futures and Peanut futures, aiming to provide a comprehensive analysis of their interconnected dynamics within the market.

In exploring the relationship between Peanut futures and other futures, the analysis can be approached from two perspectives: price and logarithmic returns. Additionally, the relationship can be examined through both static and dynamic frameworks. In this study, the static relationship is investigated using methods such as correlation matrix analysis, cointegration analysis, Granger causality tests, and multiple regression. For the dynamic relationship, a Vector Autoregressive (VAR) model\citep{sims1980macroeconomics} is employed to model the interactions, with a focus on impulse response mechanisms. Furthermore, to account for the mutual influence of volatility, EGARCH models are constructed and the Dynamic Conditional Correlation Exponential Generalized Autoregressive Conditional Heteroskedasticity (DCC-EGARCH) model\citep{engle2002dynamic}\citep{nelson1991conditional} is utilized to uncover the dynamic relationships between peanut futures and other futures.

Building on the comprehensive analysis of both static and dynamic relationships, a deep learning neural network model is designed. This study compares the performance of three models—Multilayer Perceptron (MLP)\citep{rumelhart1986learning}, Convolutional Neural Network (CNN)\citep{lecun1998gradient}, and Long Short-Term Memory (LSTM)\citep{hochreiter1997long}—in predicting Peanut futures prices, with particular attention to the impact of varying time-step training data on prediction accuracy. This multifaceted approach ensures a robust and nuanced understanding of the interplay between peanut futures and other futures markets. 

The article is structured as follows: after summarizing related work in Sec.~\ref{sec:related_work}, we do some data analysis about the historical futures data (Sec.~\ref{sec:data_analsis}) including correlation analysis, cointegration analysis and Granger causality analysis. We also carry out multiple-regression for Peanut futures (Sec.~\ref{sec:mul_regression}). We show dynamic correlation analysis in Sec.~\ref{sec:data_dynamic}. Lastly, we discuss limitations and conclusions in Secs.~\ref{sec:limitations}~and~\ref{sec:conclusions}, respectively.

\section{Related Work}
\label{sec:related_work}
The exploration of Peanut futures has emerged as a focal point in contemporary financial research, particularly following the introduction of Peanut futures contracts on the Zhengzhou Commodity Exchange (ZCE) in 2021. This burgeoning interest has spurred a lot of studies examining diverse facets of Peanut futures, encompassing price volatility, market efficiency, risk management, and intermarket relationships. 

\section{Data Analysis}

\label{sec:data_analsis}

\subsection{Data Statistics}
\label{sec:data_stat}

\begin{figure*}[t!]
    \centering
    \includegraphics[width=.9\textwidth]{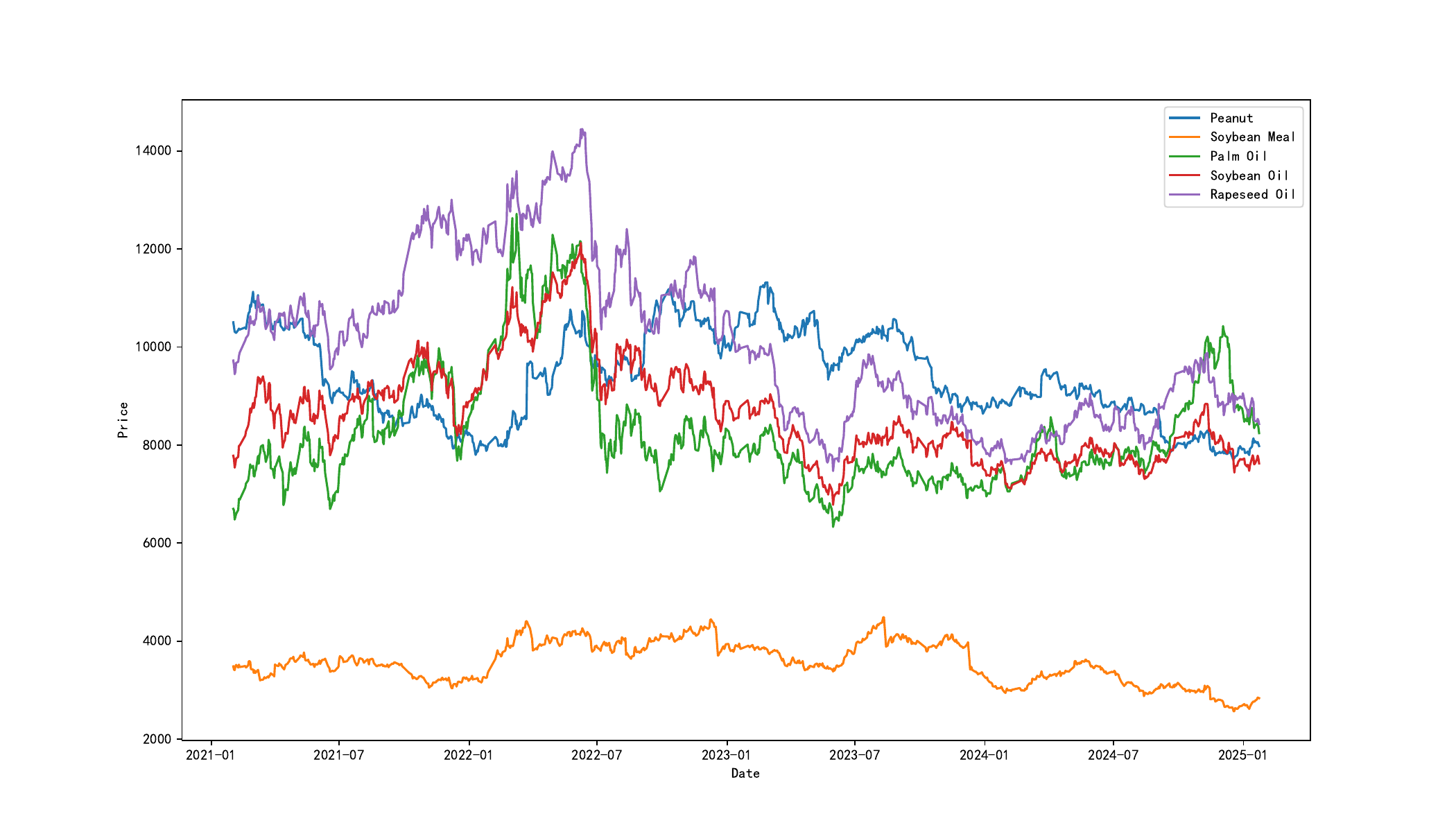}
    
    \caption{
       Five dominant contract futures price series (Peanut, Soybean Meal, Palm Oil, Soybean Oil, Rapeseed Oil)
    }
    \label{fig:price}
\end{figure*}

Figure 1 illustrates long-term trajectories of several key agricultural commodities, including Peanut, Soybean Meal, Palm Oil, Soybean Oil, and Rapeseed Oil, spanning a four-year period from January 2021 to January 2025. The price of peanut futures peaked in 2022 and then entered a bear market. Other related futures varieties also showed similar trends. Figure 1 also provides a comprehensive view of both short-term fluctuations and in commodity prices. We consider use these historical data to carry out  a comparative analysis, highlighting potential correlations or divergences in their price movements over time. Traditional time series techniques are instrumental in identifying seasonal patterns, assessing market volatility, and evaluating the impact of external factors such as geopolitical events or supply chain disruptions on commodity markets. We hope to understand the dynamics of agricultural pricing and to derive insights into market behavior. This paper attempts to uncover the static and dynamic correlations between several futures prices, thereby providing insights for analysis and forecasting.

\begin{figure*}[t!]
    \centering
    \includegraphics[width=.9\textwidth]{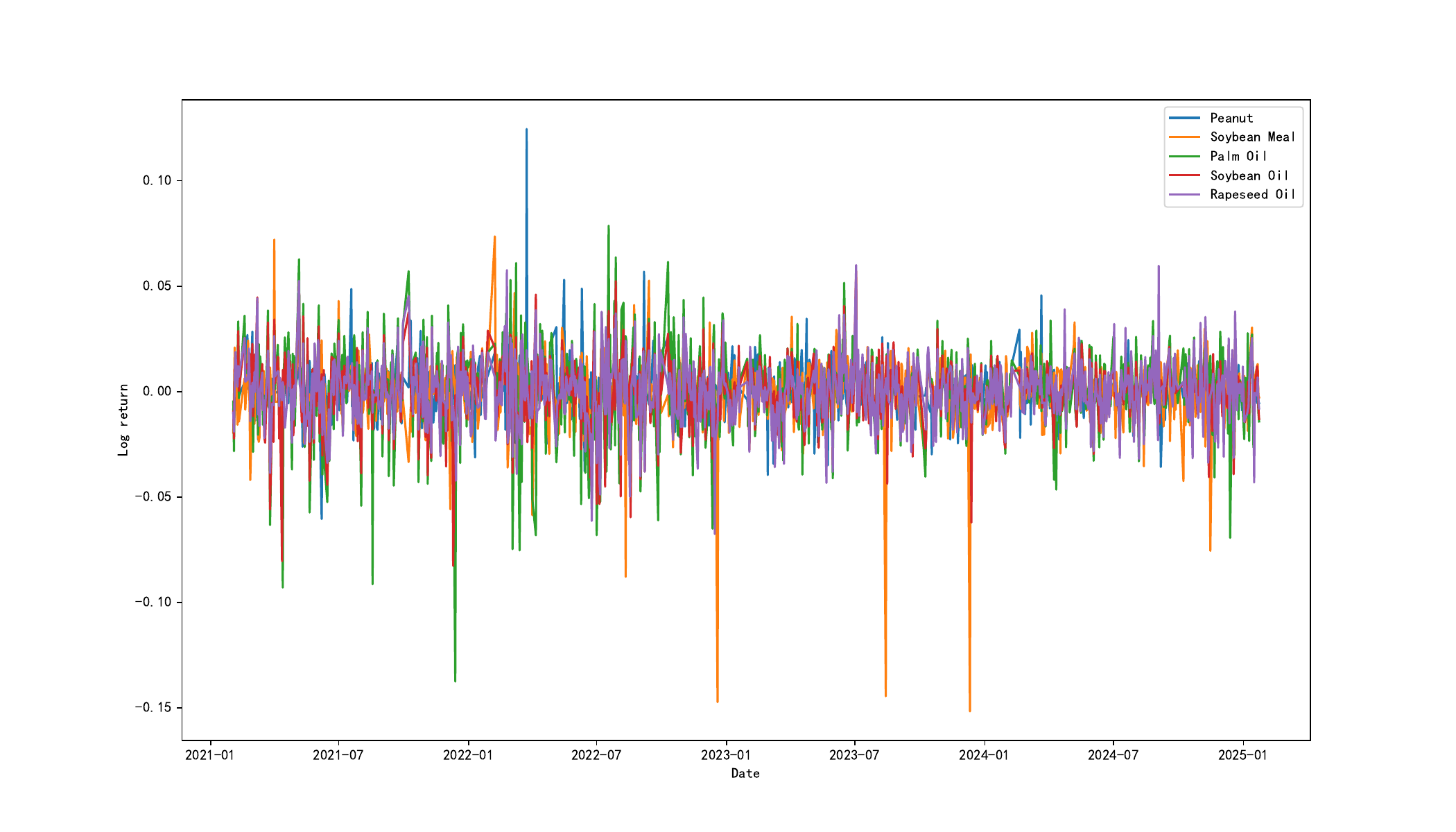}
    
    \caption{
       The logarithmic returns of five  dominant contract futures price series (Peanut, Soybean Meal, Palm Oil, Soybean Oil, Rapeseed Oil)
    }
    \label{fig:logreturn}
\end{figure*}

\paragraph{Figure 2 presents the logarithmic returns}
\label{sec:logreturn}

of five commodities (Peanut, Soybean Meal, Palm Oil, Soybean Oil, and Rapeseed Oil) over a time series spanning from January 2021 to January 2025. The logarithmic returns, which are a transformation of price data used to normalize returns and facilitate the analysis of relative price changes over time. The values range from -0.15 to 0.10, indicating both positive and negative returns, while the delineates the timeline at six-month intervals.

Logarithmic return is an important tool in financial analysis, as it more accurately reflects the continuous changes in asset prices.  They provide a more stable variance and are additive over time, making them suitable for modeling and forecasting. Single-period logarithmic return is the change in asset price from time $t-1$ to time $t$, the logarithmic return is calculated as:
\[
r_t = \ln\left(\frac{P_t}{P_{t-1}}\right)
\]
where \( P_t \) is the asset price at time \( t \), \( P_{t-1} \) is the asset price at time \( t-1 \), \( \ln \) is the natural logarithm function.

From Figure 2 we can examine the stationarity and volatility of the time series data. As can be seen from Figure 2, the five time series are generally stationary because they oscillate around zero, and their variances do not change over time.

\begin{figure*}[t!]
    \centering
    \includegraphics[width=.9\textwidth]{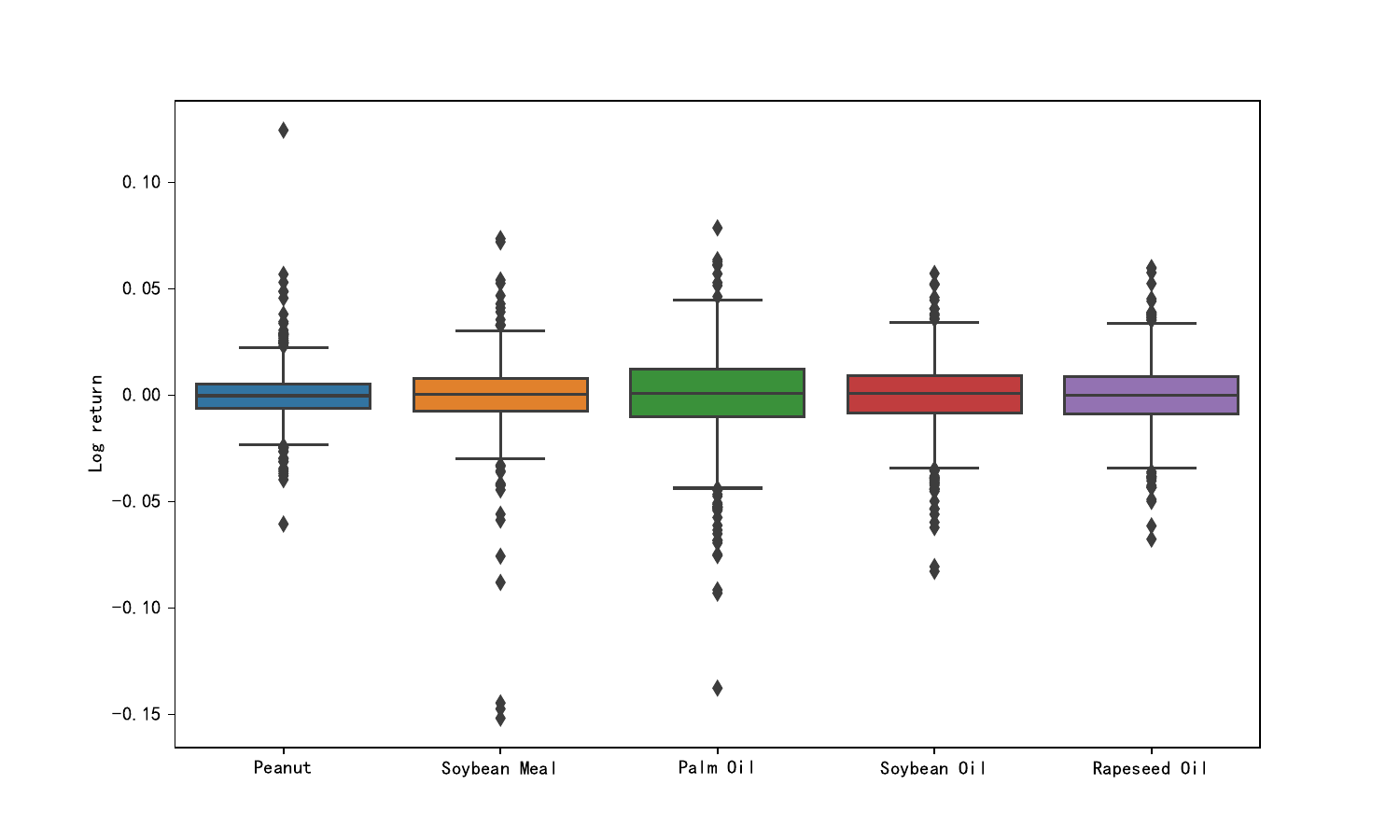}
    
    \caption{
     Box plot of logarithmic returns
    }
    \label{fig:logretbox}
\end{figure*}

The Table 1 presents the statistical analysis results of logarithmic returns for five different commodities: Peanut, Soybean Meal, Palm Oil, Soybean Oil, and Rapeseed Oil. The analysis includes key metrics such as mean, variance, skewness, and kurtosis, which describe the distribution characteristics of the returns. Additionally, the results of the Jarque-Bera (JB) normality test indicate that none of the commodities follow a normal distribution (all marked as "No"). Furthermore, the Augmented Dickey-Fuller (ADF) test p-values are all 0.0, confirming that the logarithmic returns for all commodities are weakly stationary (all marked as "Yes"). This suggests that the return series do not exhibit unit roots and are suitable for further time series analysis.

\begin{table}[htbp]
\centering
\caption{Statistical Analysis Results of Logarithmic Returns}
\begin{tabular}{lccccc}
\toprule
 & \textbf{Mean} & \textbf{Variance} & \textbf{Skewness} & \textbf{Kurtosis} & \textbf{Normality Test (JB)} \\
\midrule
Peanut & -0.000285 & 0.000138 & 1.420725 & 15.33581 & No \\
Soybean Meal & -0.000214 & 0.000249 & -2.641546 & 25.397422 & No \\
Palm Oil & 0.000215 & 0.000413 & -0.774423 & 3.821658 & No \\
Soybean Oil & -0.000021 & 0.000238 & -0.61553 & 2.669902 & No \\
Rapeseed Oil & -0.000148 & 0.000224 & -0.086483 & 1.589099 & No \\
\midrule
 & \textbf{ADF p-value} & \textbf{Weak Stationary} \\
\midrule
Peanut & 0.0 & Yes \\
Soybean Meal & 0.0 & Yes \\
Palm Oil & 0.0 & Yes \\
Soybean Oil & 0.0 & Yes \\
Rapeseed Oil & 0.0 & Yes \\
\bottomrule
\end{tabular}
\label{tab:stats}
\end{table}

\paragraph{The box plot presented in Figure 3 illustrates}
\label{sec:data_box}
the distribution of logarithmic returns for several commodities. From Figure 3, we can observe several characteristics such as central tendency,
dispersion, skewness. A box plot is crucial for identifying differences in the risk and return profiles of the commodities. It also aid in detecting anomalies or extreme values that could signify significant market events or data errors. For example, Figure 3 shows that palm oil has the largest fluctuation range, while peanuts have a relatively smaller fluctuation range. Peanuts exhibit more anomalous data compared to other futures, and the outliers are also larger. Due to the low trading volume of peanut futures and the limited participation of major players, it is inferred that their prices are more susceptible to manipulation and significant fluctuations. On the other hand, soybean meal futures have recently seen a significant increase in open interest, with a large number of participants and trading volume, making it a major market variety among agricultural products.

\begin{figure*}[t!]
    \centering
    \includegraphics[width=.45\textwidth]{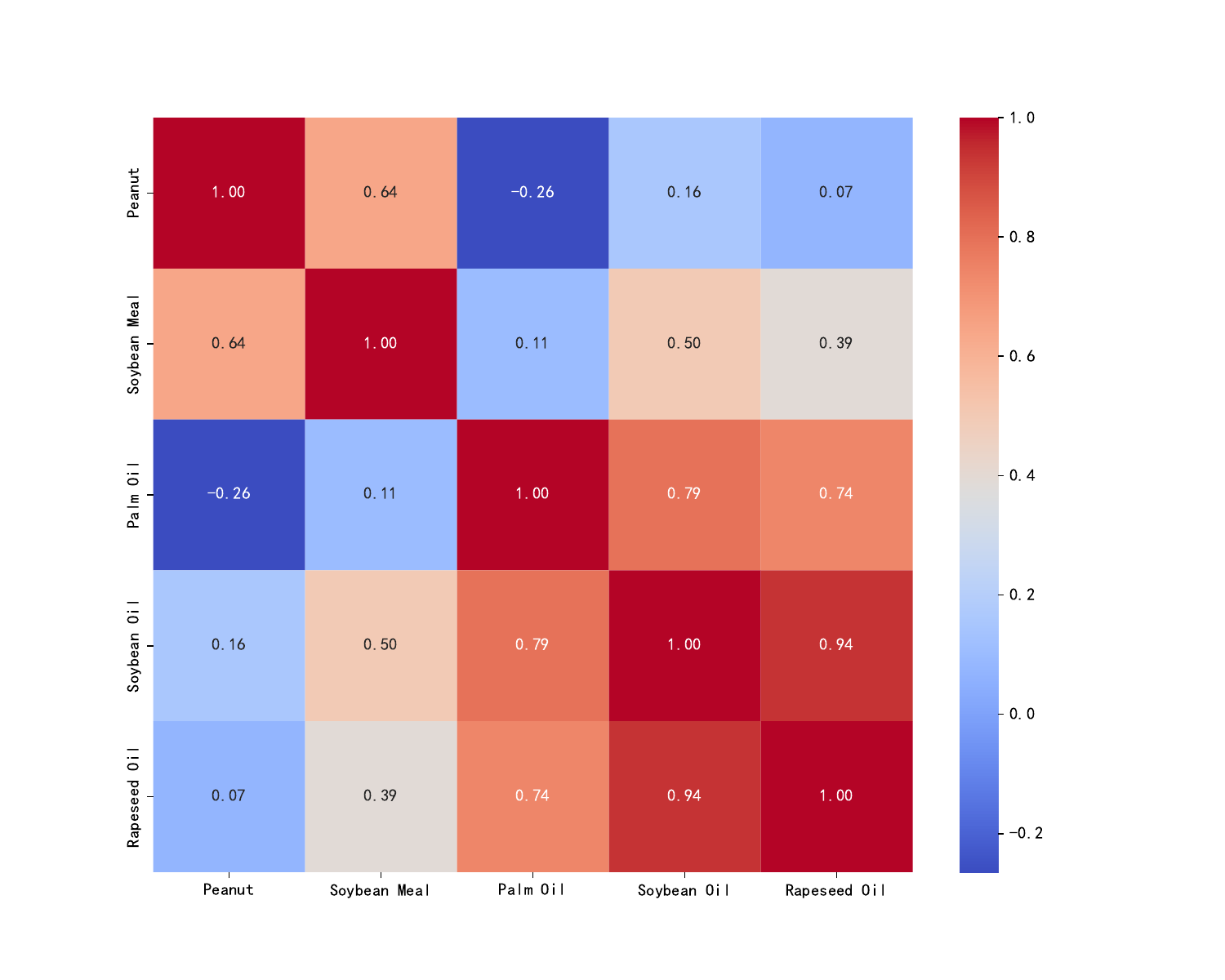}
    \quad
    \includegraphics[width=.45\textwidth]{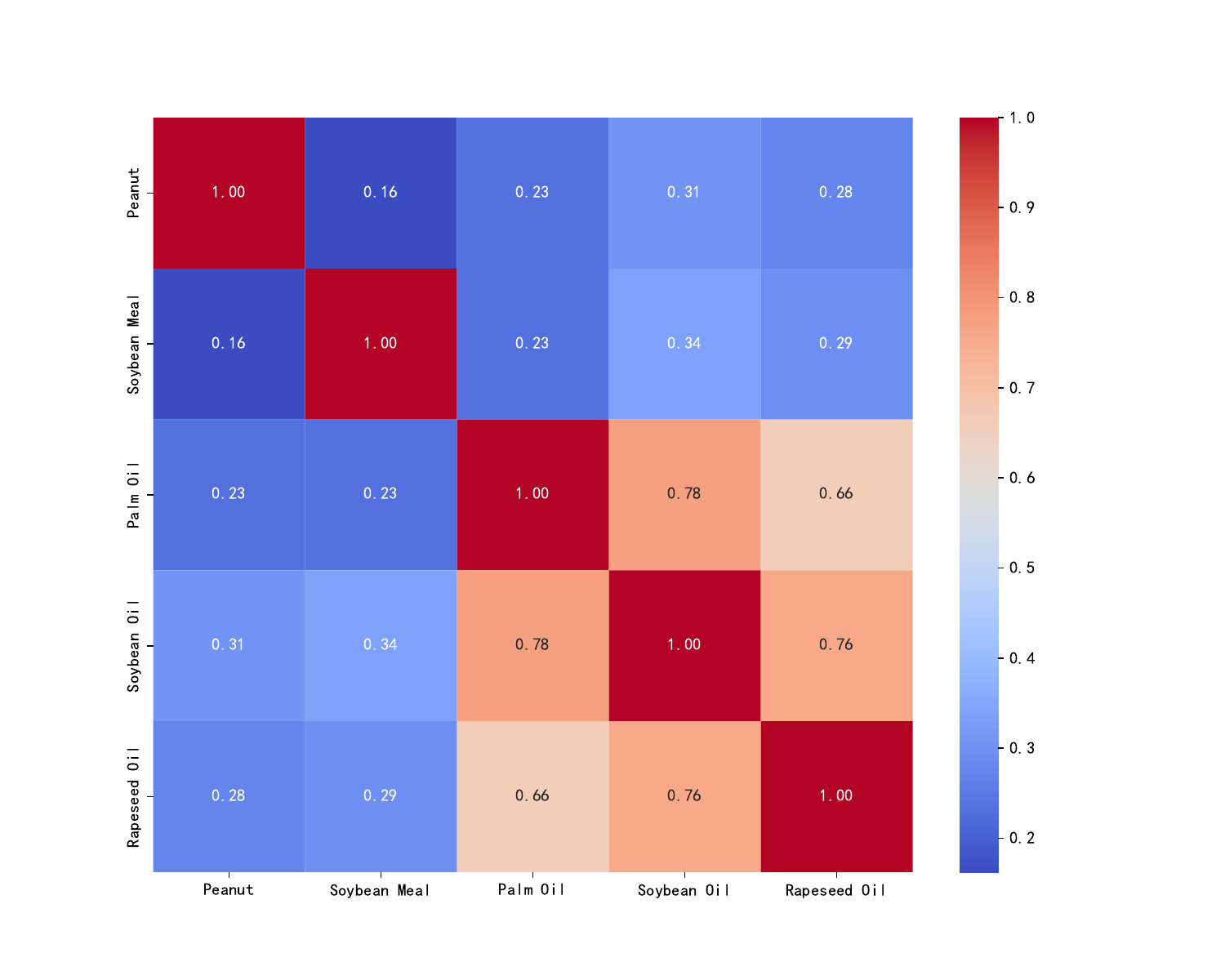}
    \caption{
     Prices correlation heatmap (left) and  logarithmic returns correlation heatmap
    }
    \label{fig:bias_modeling}
\end{figure*}

\paragraph{The correlation heatmap depicted in Figure 4 provides} a comprehensive visualization of the pairwise correlations among the future prices of five commodities: Peanut, Soybean Meal, Palm Oil, Soybean Oil, and Rapeseed Oil. The heatmap employs a color gradient to represent correlation coefficients ranging from -0.2 to 1.0, where darker shades signify stronger positive correlations and lighter shades indicate weaker or negative relationships. Notably, Soybean Oil and Rapeseed Oil exhibit a remarkably high correlation coefficient of 0.94, suggesting a strong positive interdependence in their price movements, likely driven by similar market dynamics or substitutability. Palm Oil also demonstrates significant positive correlations with Soybean Oil (0.79) and Rapeseed Oil (0.74), further underscoring the interconnectedness within this group of commodities. In contrast, Peanut shows a weak negative correlation with Palm Oil (-0.26), indicating divergent price behaviors that may offer diversification benefits in portfolio construction. The moderate correlations between Soybean Meal and Peanut (0.64) as well as Soybean Meal and Soybean Oil (0.50) reflect partial co-movements, potentially influenced by overlapping supply chains or demand factors. From the perspective of logarithmic returns, we ranked the correlation strength of peanuts from strongest to weakest as follows: soybean meal, soybean oil, rapeseed oil, and palm oil. However, peanuts exhibit a negative correlation with palm oil. Figure 4 not only highlights the varying degrees of price interdependence among the five commodities but also serves as a critical tool for identifying diversification opportunities and understanding the underlying market structures. Such insights are invaluable for developing robust risk management strategies and predictive models for peanut futures.

\paragraph{The Figure 5 presents the p-value matrix }

\begin{figure*}[t!]
    \centering
    \includegraphics[width=.9\textwidth]{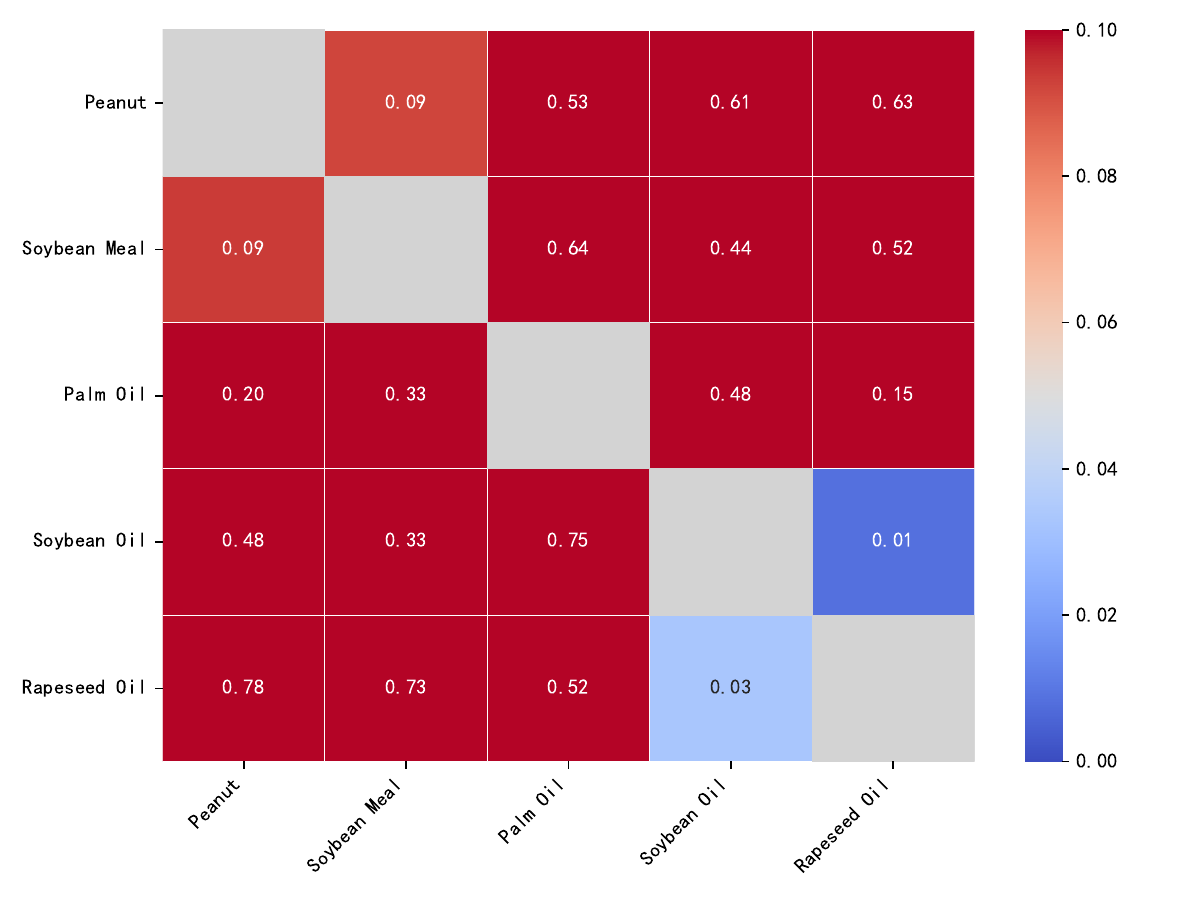}
    
    \caption{
     P-value matrix of cointegration test
    }
    \label{fig:cointegration}
\end{figure*}

derived from a cointegration test conducted on the futures price series of five commodities: Peanut, Soybean Meal, Palm Oil, Soybean Oil, and Rapeseed Oil. Cointegration tests are employed to determine whether a long-term equilibrium relationship exists between pairs of non-stationary time series, which is crucial for understanding the interdependencies and potential predictability of their price movements. The p-values in the matrix indicate the significance levels of the cointegration relationships between each pair of commodities. A p-value below the conventional threshold of $0.05$ suggests the presence of a statistically significant cointegrating relationship. We can observe from matrix that the p-value between Soybean Oil and Rapeseed Oil is $0.03$, indicating a significant cointegrating relationship at the $5\%$ level. This suggests that these two commodities share a long-term equilibrium relationship, likely driven by similar market dynamics or substitutability. The p-value between Palm Oil and Rapeseed Oil is $0.04$, also indicating a significant cointegrating relationship. Most other pairs exhibit p-values well above the $0.05$ threshold, such as Peanut and Soybean Meal (p = 0.09), Peanut and Palm Oil (p = 0.53), and Soybean Meal and Soybean Oil (p = 0.44). These results suggest the absence of a long-term equilibrium relationship between these pairs. Because we cannot find long-term equilibrium price movements between peanut and other commodities, we should consider alternative approaches when analyzing their price dynamics.

\paragraph{The Figure 6 presents a pairwise Granger causality p-value matrix}
\begin{figure*}[t!]
    \centering
    \includegraphics[width=.9\textwidth]{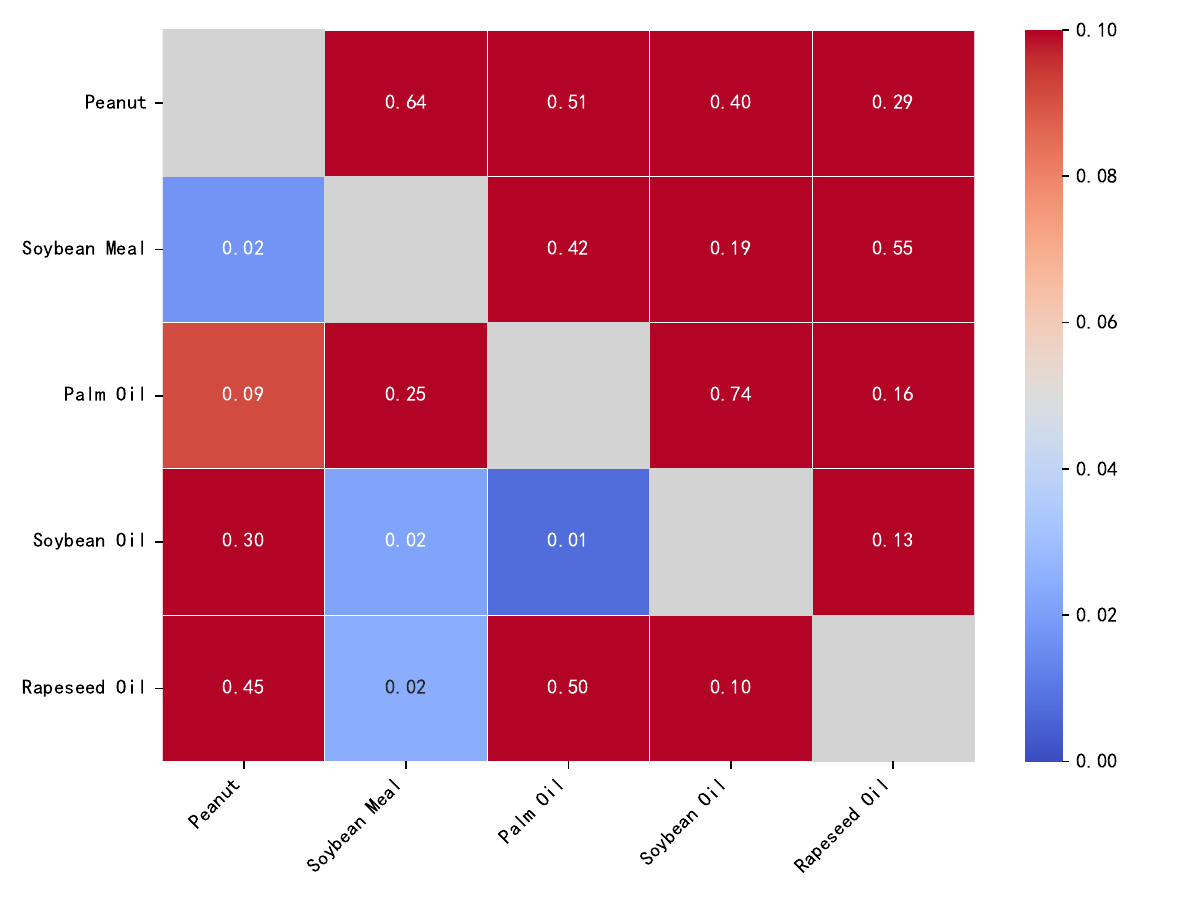}
    
    \caption{
     P-value matrix of Granger causality test
    }
    \label{fig:granger}
\end{figure*}
among five agricultural commodities: Peanut, Soybean Meal, Palm Oil, Soybean Oil, and Rapeseed Oil. Each cell indicates the p-value testing whether the row variable Granger causes the column variable. Redundant diagonal elements are omitted, as self-causality is not applicable. Significant causal relationships (p $<$ 0.05) are observed in multiple directions. For instance, Soybean Meal exhibits strong predictive power over Peanut (p = 0.02) suggesting its role as a leading indicator in predicting Peanut. Similarly, Soybean Oil demonstrates significant Granger causality toward Palm Oil (p = 0.01), reflecting potential substitution effects or shared supply chain influences. Conversely, Peanut shows no statistically significant causal relationships with other commodities (p $>$ 0.10 for all pairs), indicating its price dynamics may be driven by idiosyncratic factors.

Notably, Rapeseed Oil Granger causes Soybean Meal (p = 0.02), hinting at feedback mechanisms within the oilseed complex. However, bidirectional insignificance persists between Palm Oil and Rapeseed Oil (p = 0.74 and p = 0.50), underscoring their decoupled price behaviors. The results collectively reveal a hierarchical causal structure, with Soybean Meal and Soybean Oil acting as central nodes, while Peanut remains peripheral.

\subsection{Multiple-Regression for Peanut Futures}
\label{sec:mul_regression}

The multivariate regression model with Peanut as the dependent variable and Soybean Meal, Palm Oil, Soybean Oil, and Rapeseed Oil as independent variables can be formally expressed as:  

\[
\text{Peanut}_t = \beta_0 + \beta_1 \text{Soybean Meal}_t + \beta_2 \text{Palm Oil}_t + \beta_3 \text{Soybean Oil}_t + \beta_4 \text{Rapeseed Oil}_t + \epsilon_t
\]  

\(\text{Peanut}_t\): Price or return of the Peanut futures at time \(t\). \(\text{Soybean Meal}_t, \text{Palm Oil}_t, \text{Soybean Oil}_t, \text{Rapeseed Oil}_t\): Prices or returns of the respective futures at time \(t\). \(\beta_0\): Intercept term. \(\beta_1, \beta_2, \beta_3, \beta_4\): Regression coefficients quantifying the marginal effect of each independent variable on Peanut. \(\epsilon_t\): Error term at time \(t\), assumed to be independently and identically distributed (i.i.d.).  

This model investigates how contemporaneous movements in the four oilseed-related futures collectively explain variations in Peanut prices, controlling for interdependencies identified in the Granger causality analysis. The ordinary least squares (OLS) regression model investigates the relationship between Peanut futures prices (dependent variable) and four independent variables: Soybean Meal, Palm Oil, Soybean Oil, and Rapeseed Oil. The model achieves an R-squared value of 0.575, indicating that approximately 57.5\% of the variance in Peanut prices is explained by the included predictors. The adjusted R-squared (0.573) closely aligns with the unadjusted value, suggesting minimal overfitting. A statistically significant F-statistic (\(F = 324.7, \, p < 0.001\)) confirms the joint significance of all independent variables. Detail information is in Table 2.

Among the predictors, Soybean Meal exhibits a strong positive association with Peanut prices (\(\beta = 0.938, \, p < 0.001\)), implying that a one-unit increase in Soybean Meal futures corresponds to a 0.938-unit rise in Peanut prices, holding other variables constant. Similarly, Soybean Oil demonstrates a significant positive effect (\(\beta = 0.746, \, p < 0.001\)). Conversely, Palm Oil (\(\beta = -0.533, \, p < 0.001\)) and Rapeseed Oil (\(\beta = -0.188, \, p < 0.001\)) show statistically significant negative relationships with Peanut prices. The intercept term (\(\beta_0 = 5907.11, \, p < 0.001\)) reflects the baseline price of Peanut when all independent variables are zero.  

Diagnostic tests reveal potential issues. The Durbin-Watson statistic (0.050) signals strong positive autocorrelation in residuals, violating the independence assumption. The Jarque-Bera test (\(p = 0.029\)) rejects the null hypothesis of normally distributed residuals, indicating non-normality. These findings suggest caution in interpreting standard errors, though the large sample size (\(N = 965\)) may mitigate some biases.

\begin{table}[htbp]
  \centering
  \caption{OLS Regression Results for Peanut Futures Prices}
  \label{tab:simple_regression}
  \begin{tabular}{|l|r|r|r|r|r|}
    \hline
    \textbf{Variable} & \textbf{Coeff.} & \textbf{Std. Err.} & \textbf{t-stat} & \textbf{p-value} & \textbf{95\% CI} \\
    \hline
    Constant         & 5907.11$^{***}$ & 210.57            & 28.05           & 0.000            & [5493.9, 6320.3] \\
    Soybean Meal     & 0.938$^{***}$   & 0.067             & 13.91           & 0.000            & [0.806, 1.070]   \\
    Palm Oil         & -0.533$^{***}$  & 0.032             & -16.89          & 0.000            & [-0.595, -0.471] \\
    Soybean Oil      & 0.746$^{***}$   & 0.078             & 9.62            & 0.000            & [0.594, 0.898]   \\
    Rapeseed Oil     & -0.188$^{***}$  & 0.035             & -5.32           & 0.000            & [-0.257, -0.118] \\
    \hline
  \end{tabular}
  
  \vspace{0.1cm}
  \footnotesize
  Notes: $N = 965$, $R^2 = 0.575$, Adj. $R^2 = 0.573$, F-stat = 324.7 (p $<$ 0.001). \\
  Significance: $^{***}$ p $<$ 0.01. Standard errors are non-robust.
\end{table}

The results highlight complex interdependencies among agricultural commodities. Soybean Meal and Soybean Oil act as complementary drivers of Peanut prices, possibly due to shared demand in animal feed or biofuel sectors. Conversely, Palm Oil and Rapeseed Oil exhibit substitutive effects, potentially reflecting competitive market dynamics. While the model demonstrates strong explanatory power, residual autocorrelation and non-normality warrant further investigation, such as incorporating lagged terms or employing robust standard errors.

This paper also conducts multiple regression on the logarithmic returns of each futures price. Table 3 shows the results.
\begin{table}[htbp]
  \centering
  \caption{OLS Regression Results for logarithmic returns of Peanut Futures}
  \label{tab:simple_regression2}
  \begin{tabular}{|l|r|r|r|r|r|}
    \hline
    \textbf{Variable} & \textbf{Coeff.} & \textbf{Std.Err.} & \textbf{t-stat} & \textbf{p-value} & \textbf{95\% CI} \\
    \hline
    Constant         & -0.0003         & 0.000            & -0.71           & 0.476            & [-0.001, 0.000]  \\
    Soybean Meal     & 0.0448          & 0.024            & 1.84            & 0.066            & [-0.003, 0.093]  \\
    Palm Oil         & -0.0186         & 0.029            & -0.65           & 0.516            & [-0.075, 0.037]  \\
    Soybean Oil      & 0.1823$^{***}$  & 0.044            & 4.12            & 0.000            & [0.095, 0.269]   \\
    Rapeseed Oil     & 0.0774$^{*}$    & 0.038            & 2.06            & 0.040            & [0.004, 0.151]   \\
    \hline
  \end{tabular}
  
  \vspace{0.2cm}
  \footnotesize
  Notes: \( N = 964 \), \( R^2 = 0.103 \), Adj. \( R^2 = 0.100 \), \( F(4, 959) = 27.63 \) (\( p < 0.001 \)). \\
  Significance: \( ^{***} p < 0.01 \), \( ^{*} p < 0.05 \). Standard errors are non-robust.
\end{table}

The model achieves a low $R^2$ of 0.103, indicating that only 10.3\% of the variance in Peanut prices is explained by the predictors. The statistically significant F-statistic (\(F = 27.63, \, p < 0.001\)) suggests that the model as a whole has predictive power, though the weak explanatory capacity implies limited practical relevance. Among the independent variables, Soybean Oil exhibits a strong positive effect (\(\beta = 0.182, \, p < 0.001\)), implying that a one-unit increase in Soybean Oil futures corresponds to a 0.182-unit rise in Peanut prices, ceteris paribus. Rapeseed Oil also shows a marginally significant positive association (\(\beta = 0.077, \, p = 0.040\)). In contrast, Soybean Meal (\(\beta = 0.045, \, p = 0.066\)) and Palm Oil (\(\beta = -0.019, \, p = 0.516\)) are statistically insignificant at the 5\% level, with the former approaching marginal significance. The intercept term (\(\beta_0 = -0.0003, \, p = 0.476\)) is negligible and statistically insignificant.  

Residual diagnostics raise concerns about model assumptions. The Jarque-Bera test (\(JB = 13330.98, \, p < 0.001\)) and Omnibus test (\(p < 0.001\)) strongly reject the null hypothesis of normality, indicating severe non-normality in residuals, likely driven by extreme skewness (\(\text{Skew} = 1.55\)) and kurtosis (\(\text{Kurtosis} = 20.95\)). While the Durbin-Watson statistic (\(2.056\)) suggests no significant autocorrelation, the non-normal residuals undermine the reliability of standard inference procedures. The model identifies Soybean Oil as the primary driver of Peanut price movements, with weaker contributions from Rapeseed Oil. Low explanatory power (\(R^2 = 0.103\)) highlights the need to incorporate additional variables (e.g., macroeconomic factors, weather data) to better capture Peanut price dynamics. Severe residual non-normality necessitates robustness checks, such as robust standard errors or nonparametric methods, to validate coefficient significance.  

The regression analysis underscores the complexity of agricultural commodity interdependencies and the limitations of relying solely on oilseed-related futures for explaining Peanut price variability.

\subsection{Dynamic Correlation Analysis}
\label{sec:data_dynamic}
\paragraph{We first leverage Vector Autoregression (VAR) Model Analysis for our data.} The VAR(1) model is formulated as follows for a system of five agricultural commodities' logarithmic returns of futures prices:  
\[
\mathbf{y}_t = \mathbf{b} + \mathbf{B}_1 \mathbf{y}_{t-1} + \mathbf{\varepsilon}_t,
\]  
where \(\mathbf{y}_t = [\text{Peanut}_t, \text{Soybean Meal}_t, \text{Palm Oil}_t, \text{Soybean Oil}_t, \text{Rapeseed Oil}_t]'\) is the vector of endogenous variables, \(\mathbf{b}\) is a constant vector, \(\mathbf{B}_1\) is the coefficient matrix for the first lag, and \(\mathbf{\varepsilon}_t\) is the error vector with \(\mathbb{E}[\mathbf{\varepsilon}_t] = \mathbf{0}\) and covariance matrix \(\mathbf{\Sigma}\).  

For clarity, the system is expressed as:  
\[
\begin{aligned}
\text{Peanut}_t &= -0.0003 - 0.038 \text{Peanut}_{t-1} + 0.037 \text{Soybean Meal}_{t-1} \\
  &+ 0.060 \text{Palm Oil}_{t-1} - 0.080 \text{Soybean Oil}_{t-1} - 0.0001 \text{Rapeseed Oil}_{t-1} + \varepsilon_{1t}, \\
\text{Soybean Meal}_t &= -0.0002 - 0.004 \text{Peanut}_{t-1} + 0.007 \text{Soybean Meal}_{t-1} \\
  &- 0.004 \text{Palm Oil}_{t-1} - 0.069 \text{Soybean Oil}_{t-1} + 0.030 \text{Rapeseed Oil}_{t-1} + \varepsilon_{2t}, \\
\text{Palm Oil}_t &=  0.0002 -0.070\text{Peanut}_{t-1} + 0.012\text{Soybean Meal}_{t-1} \\
& -0.077\text{Palm Oil}_{t-1} + 0.237 \text{Soybean Oil}_{t-1}-0.072\text{Rapeseed Oil}_{t-1} + \varepsilon_{3t}\\
\text{Soybean Oil}_t = & -0.000008 -0.043\text{Peanut}_{t-1} + 0.001\text{Soybean Meal}_{t-1} \\
& -0.012\text{Palm Oil}_{t-1} -0.045\text{Soybean Oil}_{t-1} + 0.087\text{Rapeseed Oil}_{t-1} + \varepsilon_{4t}\\
\text{Rapeseed Oil}_t &= -0.0001 - 0.043 \text{Peanut}_{t-1} + 0.022 \text{Soybean Meal}_{t-1} \\
 &- 0.037 \text{Palm Oil}_{t-1} - 0.014 \text{Soybean Oil}_{t-1} + 0.110 \text{Rapeseed Oil}_{t-1} + \varepsilon_{5t}.
\end{aligned}
\]

In the Peanut equation dynamics, the lagged value of Palm Oil (\(\beta = 0.060, p = 0.046\)) exhibits a significant positive effect, while lagged Soybean Oil (\(\beta = -0.080, p = 0.090\)) approaches marginal significance. However, the equation for Soybean Meal demonstrates none of the lagged variables show statistically significant predictive power (\(p > 0.10\)). In the Palm Oil equation, Lagged Soybean Oil (\(\beta = 0.237, p = 0.004\)) strongly Granger-causes Palm Oil, indicating substitution or supply chain spillovers. In the Soybean Oil equation, Lagged Rapeseed Oil (\(\beta = 0.087, p = 0.094\)) weakly influences Soybean Oil prices. At last, Rapeseed Oil, its own lag (\(\beta = 0.110, p = 0.029\)) demonstrates significant persistence, suggesting autoregressive momentum.  

The Figure 7 shows the impulse response graph of the VAR model. The results show that although the lagged values of other futures have an impact on peanut futures, none of them are very significant.

\begin{figure*}[t!]
    \centering
    \includegraphics[width=.9\textwidth]{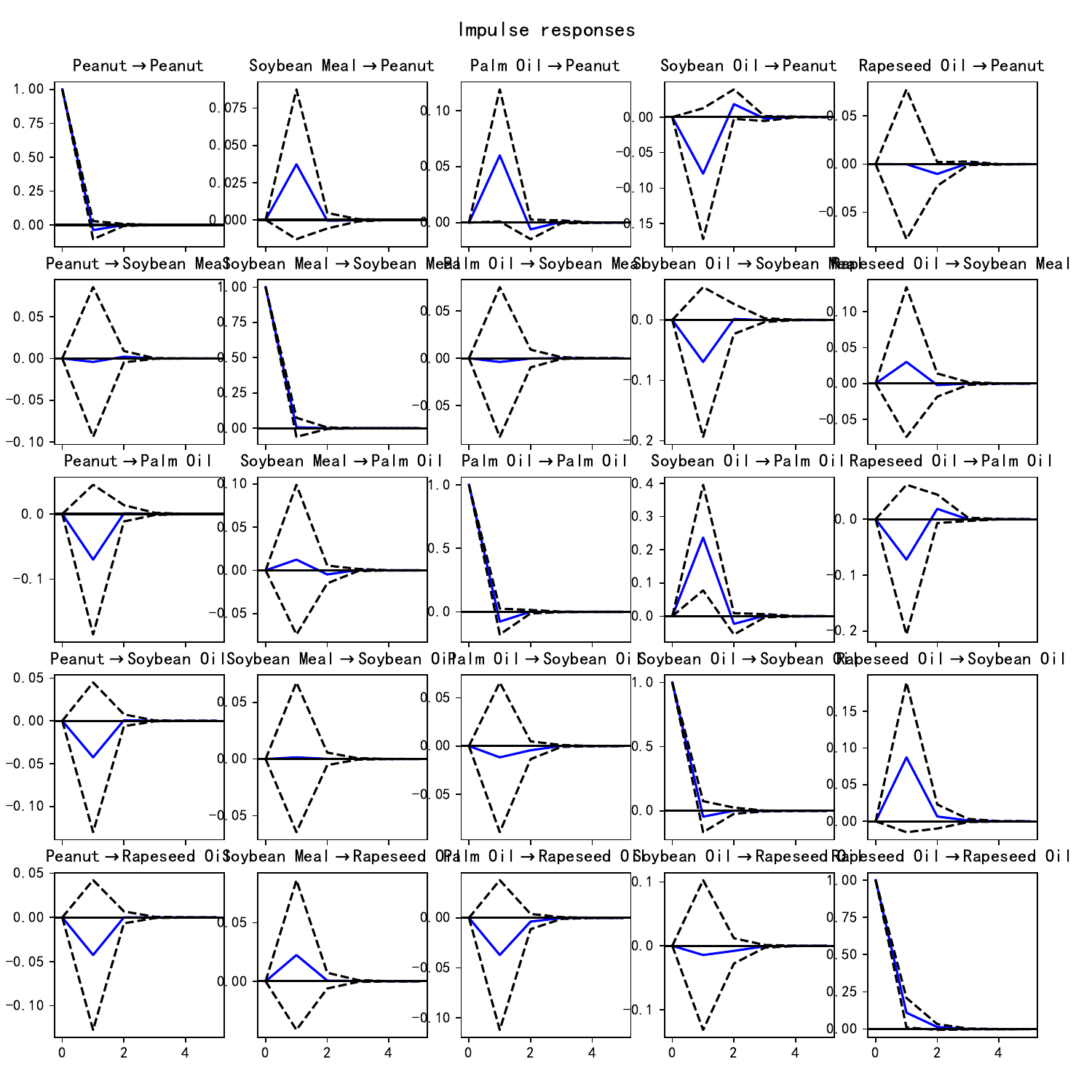}
    
    \caption{
      The impulse response graph of the VAR model
    }
    \label{fig:dcc}
\end{figure*}

\begin{figure*}[t!]
    \centering
    \includegraphics[width=.9\textwidth]{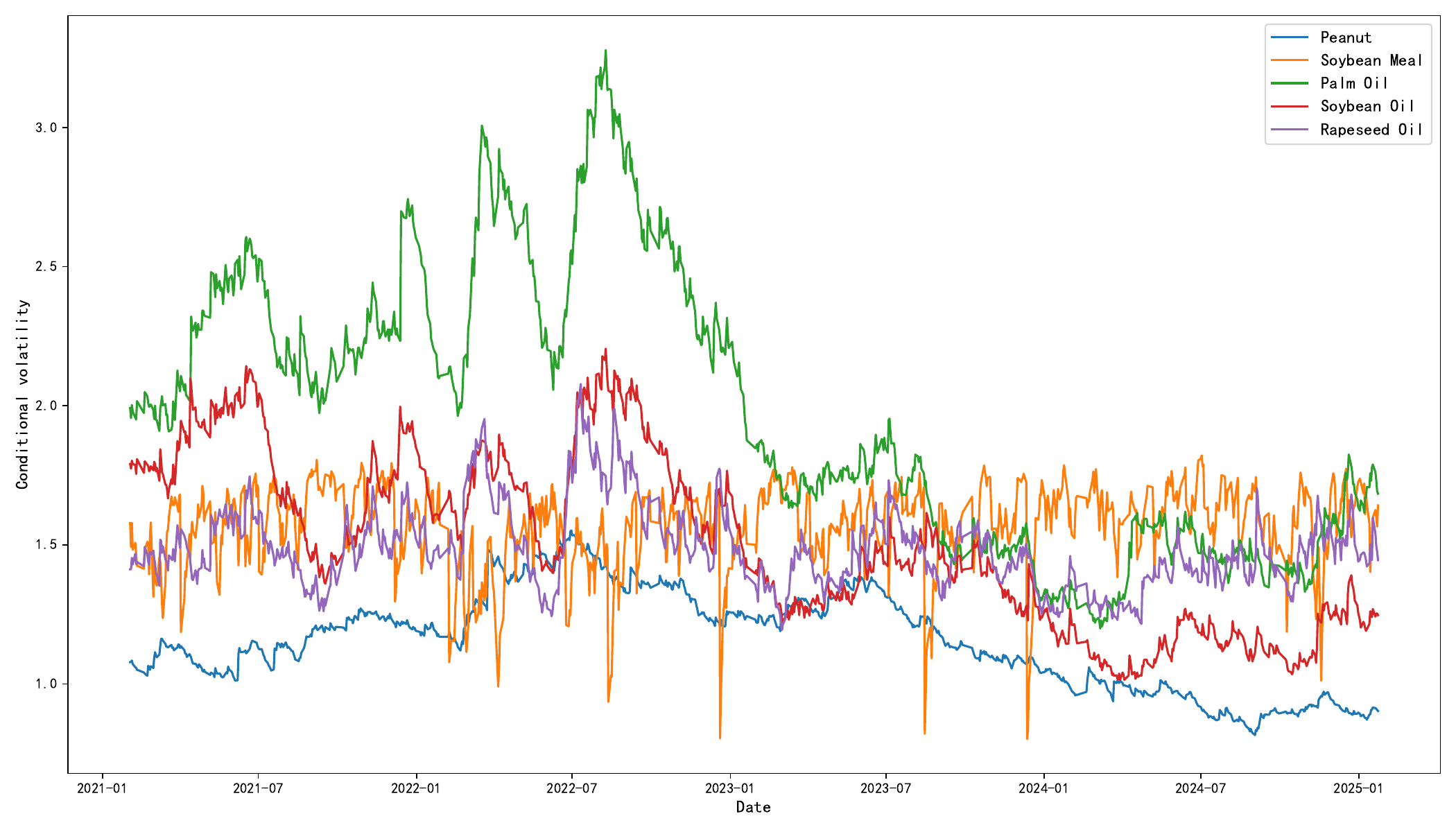}
    
    \caption{
      The conditional volatilites of the five EGARCH models 
    }
    \label{fig:vol}
\end{figure*}

\paragraph{The DCC-EGARCH model is a combination} of Dynamic Conditional Correlation (DCC) model~\citep{engle2002dynamic} and the Exponential Generalized Autoregressive Conditional Heteroskedasticity (EGARCH) model~\citep{nelson1991conditional}. It is used to model the volatility of multiple time series and their dynamic correlations while capturing asymmetric effects in volatility (e.g., the leverage effect). Due to the limitations of the GARCH model in describing Soybean Meal futures, this paper uses the EGARCH model for volatility modeling. When we modeled soy meal futures using the GARCH model, we found that the volatility was not significant, indicating that the GARCH model was not very suitable. Therefore, the EGARCH model was chosen here. For each asset \(i\), the return \(r_{i,t}\) and its conditional variance \(\sigma_{i,t}^2\) are modeled by the EGARCH(1,1) 

\[
r_{i,t} = \mu_i + \epsilon_{i,t}, \quad \epsilon_{i,t} = \sigma_{i,t} z_{i,t}
\]
where \(r_{i,t}\) is the return of asset \(i\) at time \(t\), \(\mu_i\) is the mean of asset \(i\), \(\epsilon_{i,t}\) is the residual of asset \(i\) at time \(t\), \(\sigma_{i,t}^2\) is the conditional variance of asset \(i\) at time \(t\), \(z_{i,t}\) is the standardized residual, assumed to follow a distribution with mean 0 and variance 1 (e.g., normal or Student's t-distribution). The conditional variance equation for the EGARCH(1,1) model is:

\[
\ln(\sigma_{i,t}^2) = \omega_i + \alpha_i \left( \frac{|\epsilon_{i,t-1}|}{\sigma_{i,t-1}} - \sqrt{\frac{2}{\pi}} \right) + \gamma_i \frac{\epsilon_{i,t-1}}{\sigma_{i,t-1}} + \beta_i \ln(\sigma_{i,t-1}^2)
\]
where \(\omega_i\) is the constant term, \(\alpha_i\) is the ARCH term coefficient, representing the impact of the absolute value of volatility shocks on conditional variance, \(\gamma_i\) is the asymmetry term coefficient, capturing the effect of the sign of volatility shocks on conditional variance (e.g., the "leverage effect"), \(\beta_i\) is the GARCH term coefficient, representing the impact of past conditional variance on current conditional variance.

For each asset \(i\), the standardized residual \(z_{i,t}\) is defined as

\[
z_{i,t} = \frac{\epsilon_{i,t}}{\sigma_{i,t}}
\]

The standardized residuals \(z_{i,t}\) are random variables with mean 0 and variance 1.

The DCC model describes the dynamic correlations between the standardized residuals of multiple assets~\citep{engle2002dynamic}. For \(N\) assets, the vector of standardized residuals at time \(t\) is:

\[
\mathbf{z}_t = [z_{1,t}, z_{2,t}, \dots, z_{N,t}]^T
\]

The core of the DCC model is the dynamic conditional correlation matrix \(\mathbf{R}_t\). The dynamic conditional correlation matrix \(\mathbf{R}_t\) is derived from the intermediate matrix \(\mathbf{Q}_t\):

\[
\mathbf{Q}_t = (1 - \alpha - \beta) \bar{\mathbf{Q}} + \alpha \mathbf{z}_{t-1} \mathbf{z}_{t-1}^T + \beta \mathbf{Q}_{t-1}
\]
where \(\bar{\mathbf{Q}}\) is the sample covariance matrix of the standardized residuals, defined as

  \[
  \bar{\mathbf{Q}} = \frac{1}{T} \sum_{t=1}^T \mathbf{z}_t \mathbf{z}_t^T
  \]
 \(\alpha\) and \(\beta\) are the DCC model parameters, satisfying \(\alpha \geq 0\), \(\beta \geq 0\), and \(\alpha + \beta < 1\).

The dynamic conditional correlation matrix \(\mathbf{R}_t\) is obtained by normalizing \(\mathbf{Q}_t\)

\[
\mathbf{R}_t = \text{diag}(\mathbf{Q}_t)^{-1/2} \mathbf{Q}_t \text{diag}(\mathbf{Q}_t)^{-1/2}
\]
where \(\text{diag}(\mathbf{Q}_t)\) is a diagonal matrix containing the diagonal elements of \(\mathbf{Q}_t\), \(\text{diag}(\mathbf{Q}_t)^{-1/2}\) is the inverse square root of \(\text{diag}(\mathbf{Q}_t)\). The standardized residual vector \(\mathbf{z}_t\) is assumed to follow a multivariate distribution, which can be expressed as:

\[
\mathbf{z}_t \sim \mathcal{N}(\mathbf{0}, \mathbf{R}_t) \quad \text{or} \quad \mathbf{z}_t \sim t_\nu(\mathbf{0}, \mathbf{R}_t)
\]
where  \(\mathcal{N}(\mathbf{0}, \mathbf{R}_t)\) denotes a multivariate normal distribution with mean 0 and covariance matrix \(\mathbf{R}_t\), \(t_\nu(\mathbf{0}, \mathbf{R}_t)\) denotes a multivariate Student's t-distribution with degrees of freedom \(\nu\) and covariance matrix \(\mathbf{R}_t\).

The parameters of the DCC-EGARCH model are estimated using maximum likelihood estimation (MLE)~\citep{francq2019garch}~\citep{tsay2010analysis}. The log-likelihood function is given by

\[
\mathcal{L}(\theta) = -\frac{1}{2} \sum_{t=1}^T \left( N \ln(2\pi) + \ln|\mathbf{D}_t \mathbf{R}_t \mathbf{D}_t| + \mathbf{z}_t^T \mathbf{R}_t^{-1} \mathbf{z}_t \right)
\]
where  \(\theta\) is the set of model parameters, including EGARCH and DCC parameters, \(\mathbf{D}_t\) is the diagonal matrix of conditional standard deviations, defined as

  \[
  \mathbf{D}_t = \text{diag}(\sigma_{1,t}, \sigma_{2,t}, \dots, \sigma_{N,t})
  \]

The EGARCH model can capture asymmetric effects in volatility (e.g., the "leverage effect"). The DCC model describes the dynamic conditional correlations between multiple assets. Hence the DCC-EGARCH model effectively models the volatility of multiple assets and their dynamic correlations while capturing asymmetric effects in volatility. 

Figure 8 presents the EGARCH model graphs for various commodities. From the figure, it can be observed that the volatility of Peanut futures is significantly lower than that of other futures, while Palm oil, being a popular variety in the market, exhibits higher volatility. The volatility of Soybean Meal futures is relatively stable, while the volatility of other futures varieties is generally declining.

We constructed a DCC-EGARCH model on historical data and estimated the parameters to obtain the model. The assets with the strongest dynamic correlation to peanuts, ranked by average correlation, are as follows: Soybean Oil has the highest mean correlation with peanuts at 0.306, followed by Rapeseed Oil at 0.27, Palm Oil at 0.228, and Soybean Meal at 0.166. The results of the dynamic correlation coefficients of the five futures over time are shown in the Figure 9.
\begin{figure*}[t!]
    \centering
    \includegraphics[width=.9\textwidth]{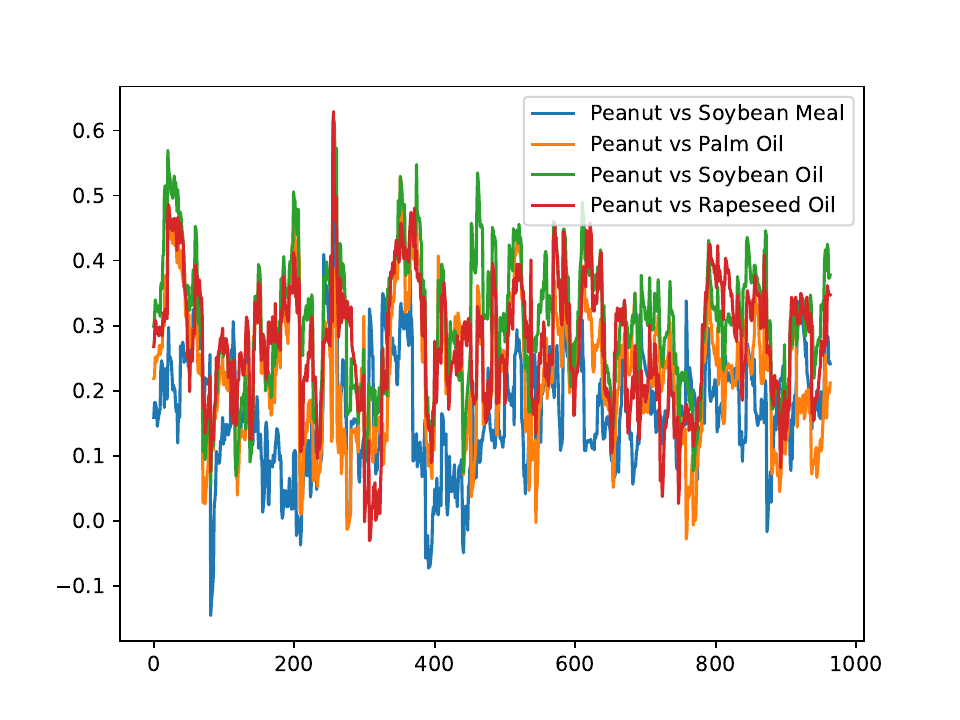}
    
    \caption{
     Dynamic conditional correlation: Peanut vs Other Futures
    }
    \label{fig:dcc}
\end{figure*}

\section{Prediction Using Neural Networks}
\label{sec:neural_network}

Although the previous sections of this paper analyzed the static and dynamic relationships between peanut futures and other agricultural futures, the predictive capability of historical data has not yet been verified. Therefore, this section implements several well-known neural networks, including MLP, CNN, and LSTM, to predict the future prices of peanuts. The lag times considered are 10 steps, 5 steps, and 1 step, with a forward prediction of 5 steps. The historical data is divided into a training dataset (80\%) and a test dataset (20\%), and the results are evaluated using MSE and MAE. In addition, this paper also considers whether to use historical data of peanuts as predictive data. Because other varieties exhibit varying degrees of correlation with peanut futures in both static and dynamic analyses.

MSE measures the average squared difference between the predicted values and the actual values. It quantifies the deviation of predictions from the true values and is more sensitive to larger errors.

\[
\text{MSE} = \frac{1}{n} \sum_{i=1}^{n} (y_i - \hat{y}_i)^2
\]

MAE measures the average absolute difference between the predicted values and the actual values. It quantifies the average absolute deviation of predictions from the true values.

\[
\text{MAE} = \frac{1}{n} \sum_{i=1}^{n} |y_i - \hat{y}_i|
\]

The scenarios are divided as follows: 4 features (excluding Peanuts) with a time step of 10; 5 features (including Peanuts) with a time step of 10; 5 features (including Peanuts) with a time step of 5; and 5 features (including peanuts) with a time step of 1. Note that the forecast horizon in all three tables is 5.

\begin{table}[htbp]
\centering
\caption{Performance with 4 features (without Peanut, time step is 10).}
\label{tab:4features}
\begin{tabular}{lccccc}
\toprule
\textbf{Model} & \textbf{Step 1} & \textbf{Step 2} & \textbf{Step 3} & \textbf{Step 4} & \textbf{Step 5} \\
\midrule
\textbf{MLP (MSE)} & 554976.65 & 553358.31 & 563178.25 & 484079.46 & 590137.62 \\
\textbf{MLP (MAE)} & 614.90 & 609.81 & 611.22 & 568.8326 & 637.41 \\
\midrule
\textbf{CNN (MSE)} & 3559658.94 & 3703819.68 & 3706808.03 & 3747304.73 & 3682506.82 \\
\textbf{CNN (MAE)} & 1508.15 & 1544.73 & 1545.73 & 1556.20 & 1537.07 \\
\midrule
\textbf{LSTM (MSE)} & 2220626.71 & 2276965.58 & 2290012.02 & 2312374.40 & 2343908.82 \\
\textbf{LSTM (MAE)} & 1411.00 & 1432.51 & 1437.06 & 1448.42 & 1461.56 \\
\bottomrule
\end{tabular}
\end{table}

\begin{table}[htbp]
\centering
\caption{Performance with 5 features (with Peanut, time step is 10).}
\label{tab:5features_timestep10}
\begin{tabular}{lccccc}
\toprule
\textbf{Model} & \textbf{Step 1} & \textbf{Step 2} & \textbf{Step 3} & \textbf{Step 4} & \textbf{Step 5} \\
\midrule
\textbf{MLP (MSE)} & 111938.37 & 200193.61 & 248546.82 & 206736.42 & 209503.49 \\
\textbf{MLP (MAE)} & 270.25 & 384.71 & 429.98 & 394.75 & 400.37 \\
\midrule
\textbf{CNN (MSE)} & 90663.54 & 84776.40 & 89771.89 & 95005.68 & 109265.06 \\
\textbf{CNN (MAE)} & 235.77 & 230.99 & 238.49 & 247.75 & 265.45 \\
\midrule
\textbf{LSTM (MSE)} & 26772.50 & 23777.70 & 28805.02 & 34235.25 & 35128.76 \\
\textbf{LSTM (MAE)} & 132.76 & 122.35 & 139.88 & 156.02 & 155.54 \\
\bottomrule
\end{tabular}
\end{table}

\begin{table}[htbp]
\centering
\caption{Performance with 5 features (with Peanut, time step is 5).}
\label{tab:5features_timestep5}
\begin{tabular}{lccccc}
\toprule
\textbf{Model} & \textbf{Step 1} & \textbf{Step 2} & \textbf{Step 3} & \textbf{Step 4} & \textbf{Step 5} \\
\midrule
\textbf{MLP (MSE)} & 115360.38 & 100540.17 & 122709.31 & 161495.37 & 141965.66 \\
\textbf{MLP (MAE)} & 258.70 & 262.68 & 265.95 & 313.93 & 303.82 \\
\midrule
\textbf{CNN (MSE)} & 48798.40 & 49423.64 & 52768.36 & 63602.18 & 63818.37 \\
\textbf{CNN (MAE)} & 170.32 & 174.59 & 183.30 & 201.25 & 202.58 \\
\midrule
\textbf{LSTM (MSE)} & 36098.45 & 63740.85 & 37076.64 & 55996.44 & 45982.56 \\
\textbf{LSTM (MAE)} & 155.59 & 204.88 & 162.81 & 191.35 & 168.83 \\
\bottomrule
\end{tabular}
\end{table}
The performance of MLP, CNN, and LSTM models was evaluated across four experiments with different feature sets and time steps. Lower values of MSE and MAE indicate better performance. Below is a summary of the results:
\begin{itemize} 

\item 4 Features (without Peanut, time step is 10) as Table 4 shows. MLP consistently achieved the lowest MSE and MAE values across all steps. For example, in Step 1, MLP achieved an MSE of 554,976.65 and an MAE of 614.90, significantly outperforming CNN and LSTM.
\item 5 Features (With Peanut, time step is 10) as Table 5 shows. LSTM demonstrated the best performance, particularly in Step 2, where it achieved the lowest MSE (23,777.70) and MAE (122.35). This trend continued across all steps, with LSTM consistently outperforming MLP and CNN.
   
\item 5 Features (With Peanut, time step is 5) as Table 6 shows. LSTM showed the best performance in this experiment, LSTM consistently achieved lower MSE and MAE values compared to MLP and CNN across all other steps except for step 2.
\item 5 Features (With Peanut, time step is 1) as Table 7 shows.  LSTM maintained its advantage almost across all steps, with CNN being a close competitor. CNN performed the best, achieving the lowest MSE (27,665.55) and MAE (131.11) in Step 1.
\end{itemize}

On limited data, perhaps we can draw some less reliable conclusions: LSTM is the best choice for longer timesteps and complex temporal dependencies; CNN is highly effective for shorter time steps and intermediate temporal complexity; MLP is a strong performer for simpler feature sets without temporal dynamics.
\begin{table}[htbp]
\centering
\caption{Performance with 5 features (with Peanut, time step is 1).}
\label{tab:5features_timestep1}
\begin{tabular}{lccccc}
\toprule
\textbf{Model} & \textbf{Step 1} & \textbf{Step 2} & \textbf{Step 3} & \textbf{Step 4} & \textbf{Step 5} \\
\midrule
\textbf{MLP (MSE)} & 57911.28 & 70843.84 & 73707.86 & 79655.15 & 65900.89 \\
\textbf{MLP (MAE)} & 188.97 & 216.92 & 224.90 & 237.06 & 204.47 \\
\midrule
\textbf{CNN (MSE)} & 27665.55 & 31355.14 & 37567.18 & 36383.37 & 46619.99 \\
\textbf{CNN (MAE)} & 131.11 & 144.79 & 160.07 & 157.64 & 177.24 \\
\midrule
\textbf{LSTM (MSE)} & 28293.46 & 29555.45 & 32412.77 & 36414.89 & 26680.28 \\
\textbf{LSTM (MAE)} & 131.23 & 142.31 & 150.01 & 154.75 & 129.51 \\
\bottomrule
\end{tabular}
\end{table}

\section{Limitations}
\label{sec:limitations}

This ariticle conducts static and dynamic analysis on limited data, and the reliability of the results is constrained. However, we attempt to uncover some valuable insights to analyze the reasons behind changes in peanut prices and logarithmic returns. Additionally, this paper does not consider other factors influencing prices and returns, which should be addressed in future research.

\section{Conclusions}
\label{sec:conclusions}
This study investigates both static and dynamic interrelationships between Peanut futures and other agricultural futures within the China's futures market. In the static analysis, a robust price correlation is identified between Peanut futures and Soybean Meal futures, while a significant Granger causality relationship is observed in their logarithmic returns. However, no substantial cointegration is found between Peanut futures and other agricultural futures, suggesting a lack of long-term equilibrium relationships. Multiple regression analysis further reveals that Peanut futures prices exhibit significant associations with other futures prices, with logarithmic returns demonstrating notable linkages to Soybean Oil and Rapeseed Oil futures. 

In the dynamic analysis, the impulse response of Peanut futures to other futures at the first lag is found to be statistically insignificant. Nonetheless, the DCC-EGARCH model highlights Soybean Oil futures as the most influential dynamic factor affecting Peanut futures, indicating pronounced volatility spillover effects. To enhance predictive capabilities, this study employs advanced neural network architectures—Multilayer Perceptron (MLP), Convolutional Neural Network (CNN), and Long Short-Term Memory (LSTM)—for Peanut price forecasting. While the neural networks demonstrate effective predictive performance, their consistency varies significantly across different training parameters, underscoring the sensitivity of these models to hyperparameter configurations. These findings contribute to a deeper understanding of the complex dynamics in agricultural futures markets and highlight the potential and limitations of machine learning techniques in financial forecasting. 

\bibliography{main}

\clearpage % Start a new page for bibliography
% \twocolumn
%\bibliographystyleapp{plainnat}
%\bibliographyapp{main} % Your bibliography

% \clearpage
\end{document}